\documentclass[12pt]{article}

\usepackage{scicite}
\usepackage[font=small,labelfont=bf]{caption}

\usepackage{times}

\usepackage{amsmath}
\usepackage{amssymb}
\usepackage{graphicx}
\usepackage{dcolumn}
\usepackage{bm}
\usepackage[rightcaption]{sidecap}
\usepackage{xcolor}

\newenvironment{sciabstract}{
\begin{quote} \bf}
{\end{quote}}

\title{\Large{Under pressure: \\Hydrogel swelling in a granular medium}}

\author{\normalsize{Jean-Fran\c cois Louf,$^{1}$ Nancy B. Lu,$^{1}$} \\\normalsize{Margaret G. O'Connell,$^{1}$ H. Jeremy Cho,$^{1,2}$ Sujit S. Datta$^{1\ast}$}\\

\normalsize{%
 $^{1}$ Department of Chemical and Biological Engineering, Princeton University, Princeton NJ 08544
}\\

\normalsize{%
 $^{2}$ Department of Mechanical Engineering, University of Nevada, Las Vegas NV 89154
}\\

\normalsize{$^\ast$To whom correspondence should be addressed; E-mail:  ssdatta@princeton.edu.}\\
}

\date{}

\begin{document}

\baselineskip24pt

\maketitle 

\begin{sciabstract}\small{Hydrogels hold promise in agriculture as reservoirs of water in dry soil, potentially alleviating the burden of irrigation. However, confinement in soil can drastically reduce the ability of hydrogels to absorb water and swell, limiting their wide-spread adoption. Unfortunately, the underlying reason remains unknown. By directly visualizing the swelling of hydrogels confined in three-dimensional (3D) granular media, we demonstrate that the extent of hydrogel swelling is determined by the competition between the force exerted by the hydrogel due to osmotic swelling and the confining force transmitted by the surrounding grains. Furthermore, the medium can itself be restructured by hydrogel swelling, as set by the balance between the osmotic swelling force, the confining force, and inter-grain friction. Together, our results provide quantitative principles to predict how hydrogels behave in confinement, potentially improving their use in agriculture as well as informing other applications such as oil recovery, construction, mechanobiology, and filtration.}\\

\small{\textbf{Teaser:} Visualization reveals that confinement in a granular medium hinders hydrogel swelling, with implications for agriculture.}


\end{sciabstract}

\newpage\section*{Introduction}

Hydrogels are cross-linked polymer networks that can absorb up to $\sim10^{3}$ times their dry weight in water while retaining integrity \cite{bertrand2016dynamics,fernandez2011microgel}. Their highly tunable sorption characteristics, biocompatibility, and ease of manufacture have made hydrogels the center of both fundamental and applied research over the past few decades \cite{wichterle1960}. As a result, hydrogels are used in many common products, such as diapers and contact lenses, as well as in emerging materials applications \cite{fernandez2011microgel,oconnell2019,yao2012,yao2015,bai2007,krafcik2017improved,dolega2017cell,lee2019dispersible,kapur1996}. A particularly promising use of hydrogels is in agriculture: dry hydrogel particles mixed into soil can absorb rain or irrigation water \cite{wei2013, wei2013_capillary_bundle_model, cejas2014, wei2014, wei2014_morphology}, acting as water reservoirs that can hydrate plants even in times of drought. Some field tests of this idea appear encouraging, with hydrogel-amended soil greatly increasing crop yield while using less water \cite{woodhouse}. Other tests, however, show less favorable results, with hydrogel amendment yielding minimal benefit to crop growth \cite{bai2010,frantz2005} and instead adversely affecting soil elastic modulus and grain packing density, potentially increasing erodibility \cite{hejduk2012,sojka1998}. The reason for this variability in outcomes remains a puzzle, and is rooted in a poor understanding of how hydrogels swell when confined in soil. Indeed, even basic characterization of this process is lacking, due to the inability to visualize hydrogels in opaque granular media. Thus, real-world uses of hydrogels proceed by trial and error---yielding highly variable results that limit the widespread adoption of hydrogels in agriculture, as well as in other applications involving their use in tight and tortuous spaces such as oil recovery, construction, mechanobiology, and filtration \cite{oconnell2019,yao2012,yao2015,bai2007,krafcik2017improved,dolega2017cell,lee2019dispersible,kapur1996}.

 Here, we report the first direct visualization of hydrogel swelling within a model 3D granular medium with tunable confining stresses and grain sizes. Our experiments enable us to measure, \textit{in situ}, two key quantities that were previously inaccessible: the extent of hydrogel swelling and medium restructuring. \textcolor{black}{Unlike an imposed osmotic or hydrostatic pressure, confinement in a granular medium subjects the surface of a hydrogel to a spatially non-uniform stress. We therefore extend the classic Flory-Rehner theory of hydrogel swelling by coupling it to Hertzian contact mechanics that explicitly treats the stresses exerted by the medium at the hydrogel-grain contacts. Using this approach, we show} that the extent of hydrogel swelling is determined by the balance between the osmotic swelling force exerted by the hydrogel and the confining force transmitted by the surrounding grains. Furthermore, we demonstrate that a balance of the same forces, also including inter-grain friction, determines the onset of restructuring of the surrounding medium. Our work therefore reveals the physical principles that describe how hydrogel swelling in and restructuring of a granular medium both depend on the properties of the hydrogel, the properties of the medium, and confining stress. \textcolor{black}{Indeed, we show that our theoretical framework describes not only our measurements, but helps to rationalize previous measurements of hydrogel water absorption in soil. Thus,} these insights expand current understanding of hydrogel swelling to more complex environments, and could ultimately be used to inform applications of hydrogels in agriculture, oil recovery, formulation of building materials, mechanobiology, and filtration.

\section*{Results}

\subsection*{Granular confinement hinders hydrogel swelling} 

\begin{figure}
     \centering
         {\includegraphics[width=0.75\textwidth]{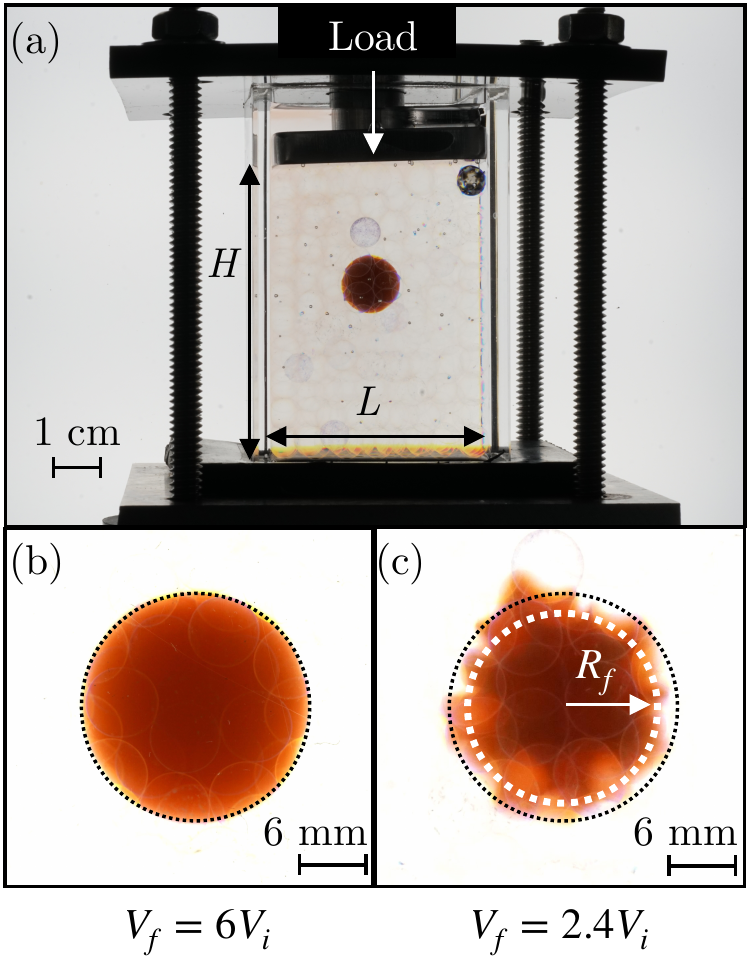}}\\
      \caption{Swelling of a hydrogel confined in a 3D granular medium. (a) Image of apparatus for testing hydrogel swelling in confinement. A hydrogel (orange) is embedded within a granular medium composed of glass beads (hazy transparent circles) packed within a transparent acrylic chamber with an overlying loaded piston. (b) Image of a hydrogel swollen within a medium in the absence of an applied load, showing that it swells freely. (c) Image of a hydrogel swollen within a medium under a strong applied load, corresponding to a confining stress of 22 kPa, showing that it deforms strongly and exhibits hindered swelling. The white dotted circle shows an inscribed circle, representing the size of the region of the hydrogel in contact with surrounding beads, for which swelling is hindered. The black dotted circle shows the outline of the hydrogel under no applied load from (b). As shown by the white space in between the black dashed outline and the projection of the hydrogel, its projected area is smaller under strong applied load. Photo Credit: Jean-Fran\c cois Louf, Princeton University.}\label{Fig1}
 \end{figure}

We prepare 3D disordered granular media by packing borosilicate glass beads with mean radii $R_{b}=1$, $1.5$, $2.5$, or $3$ mm---characteristic of coarse unconsolidated soil---$H\approx6$ cm high in a transparent, sealed, acrylic box $L\times L=4.3$ cm $\times~4.3$ cm across. While packing each medium, we place a colored polyacrylamide hydrogel sphere of initial radius $R_i\approx5.9$ mm near the center, $h\approx2$ cm from the top surface of the medium. We then repeatedly tap the container using a metal rod for $\sim20$ s; therefore, the packing approaches the random close packing limit from the initial random loose packed state, with a porosity approximately between $36$ and $41\%$. To impose a fixed confining stress, we then place a weighted piston of mass $m$ on top of the granular medium; the piston has a slight gap around its edges, enabling it to move freely along with solvent around its edges, while keeping beads confined underneath. We overfill the packing with the solvent so the liquid surface is much higher than the position of the piston. The overall apparatus with beads of a mean radius $3$ mm is shown in Figure \ref{Fig1}a. 

Light scattering from the bead surfaces typically precludes direct observation of dynamics within a granular medium. While this challenge can be overcome by infiltrating the medium with a refractive index-matched solvent \cite{krummel2013,datta2013,datta2014,dijksman2012,pouliquen2003}, these are typically poor solvents for the polymer that composes the hydrogel, and hence do not enable hydrogel swelling. We overcome both of these challenges using a 54.1 wt\% aqueous solution of NH$_{\rm{4}}$SCN, a refractive index-matched solvent that also enables hydrogel swelling. To initiate hydrogel swelling in each experiment, we completely saturate the medium with this solution; the granular medium becomes transparent, allowing direct visualization of the hydrogel using an LED light panel, as shown in Fig. \ref{Fig1}a and Movie S1. We then monitor the subsequent dynamics of the hydrogel swelling using a Nikon Micro-NIKKOR 55mm f/2.8 lens mounted on a Sony $\alpha$6300 camera, acquiring multi-color images every 30 min over a duration of 100 hours. Computer analysis of these images yields a two-dimensional (2D) projection of the overall hydrogel shape within the granular medium, characterized by a projected area $A$ and a perimeter $P$ that we directly determine from the binarized images.

\begin{figure}
     \centering
         {\includegraphics[width=0.85\textwidth]{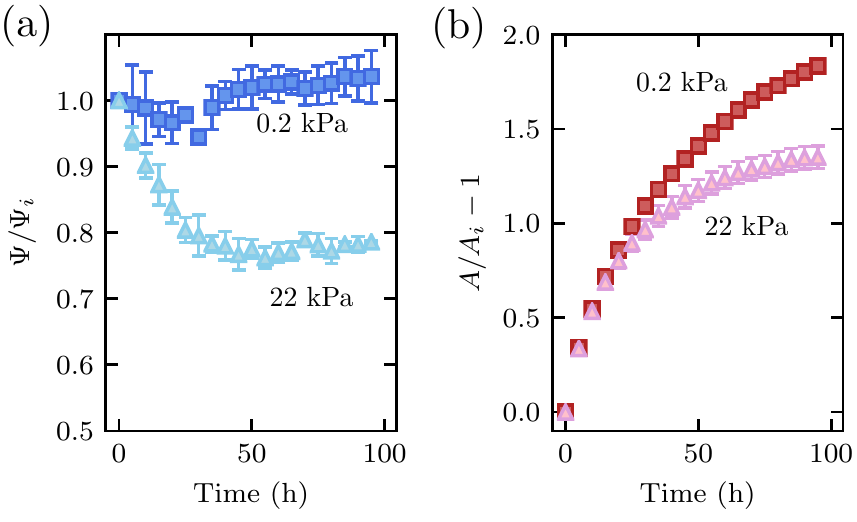}}\\
      \caption{Characterization of hydrogel swelling in confinement over time. Measurements of (a) normalized circularity $\Psi$ and (b) fractional change in projected area $A$ show that hydrogels under a small confining stress of 0.2 kPa (dark symbols) remain spherical and swell more, while hydrogels under a strong confining stress of 22 kPa (light symbols) deform into a non-spherical shape and swell less. Both quantities are normalized by their initial value. To perform these measurements, we binarize our images of hydrogel swelling inside the granular media using a given threshold value; to assess the uncertainty in these measurements, we vary this threshold value by $\pm~10\%$, for which the binarized images still closely approximate the shape of the imaged hydrogel. The resulting standard deviation in the measurements of the normalized $\Psi$ and $A$ is represented by the error bars in the plot. When not shown, error bars are smaller than the symbol size.}\label{Fig2}
 \end{figure}

In the absence of an applied load ($m=0$), the hydrogel is confined only by the weight of the overlying beads; thus, the confining stress transmitted by the surrounding beads $\sigma$ can be approximated by the gravitational stress $\Delta\rho gh\approx0.2$ kPa, where $\Delta\rho\approx1.2$ g/cm$^3$ is the density difference between the beads and the solvent and $g$ is gravitational acceleration. In this case, the hydrogel swells freely, continually rearranging the surrounding beads above it and retaining its spherical shape, as shown in Fig. \ref{Fig1}b and Movie S2.  Hence, the circularity $\Psi\equiv 4\pi A/P^{2}$ remains $\approx1$ throughout the entire swelling process as shown in Fig. \ref{Fig2}a (dark blue). The hydrogel size in turn increases over time, as reflected by the increase in the projected area $A$ shown in Fig. \ref{Fig2}b (red). The hydrogel ultimately reaches a final volume $V_{f}$ that is $\sim$ 6 times larger than its initial volume $V_i$, measured by removing it from the medium after $\sim170$ h, measuring the projected area $A$, and estimating the volume as $V\sim A^{3/2}$. 

We observe completely different swelling behavior for a hydrogel under a strong applied load with $m\approx4$ kg, corresponding to a confining stress transmitted by the individual beads $\sigma\approx\Delta\rho gh+mg/L^{2}\approx22$ kPa. In this case, the hydrogel shape changes drastically as it swells. The medium no longer rearranges, and as a result, hydrogel swelling is strongly hindered at the regions of contact with the surrounding beads. Instead, the hydrogel can only swell in between these regions of bead contact, causing it to finger into the surrounding pore space, as shown in Fig. \ref{Fig1}c and Movie S3. This fingering process does not proceed indefinitely, but plateaus after $\approx30$ h, as shown in Fig. \ref{Fig2}a (light blue). The increase in hydrogel size over time is concomitantly suppressed, as shown in Fig. \ref{Fig2}b (pink); after $\sim170$ h, the hydrogel ultimately reaches a volume $V_{f}$ that is only 2.4 times larger than its initial volume $V_i$, again measured by removing it from the medium, measuring the projected area $A$, and estimating the volume as $V\sim A^{3/2}$. To directly characterize the size of the region of hindered swelling, we determine the inscribed circle—--the largest possible circle that can be drawn inside the hydrogel, shown by the white dashed circle in Fig. \ref{Fig1}c---which isolates the region of hindered swelling without the additional influence of the hydrogel fingers that protrude in between grain contacts. The radius of this inscribed circle at the final timepoint is denoted $R_{f}$; for strong applied load, $R_{f}$ approaches $\approx R_{i}$. Clearly, confinement hinders hydrogel swelling.

\begin{figure*}
     \centering
         {\includegraphics[width=0.95\textwidth]{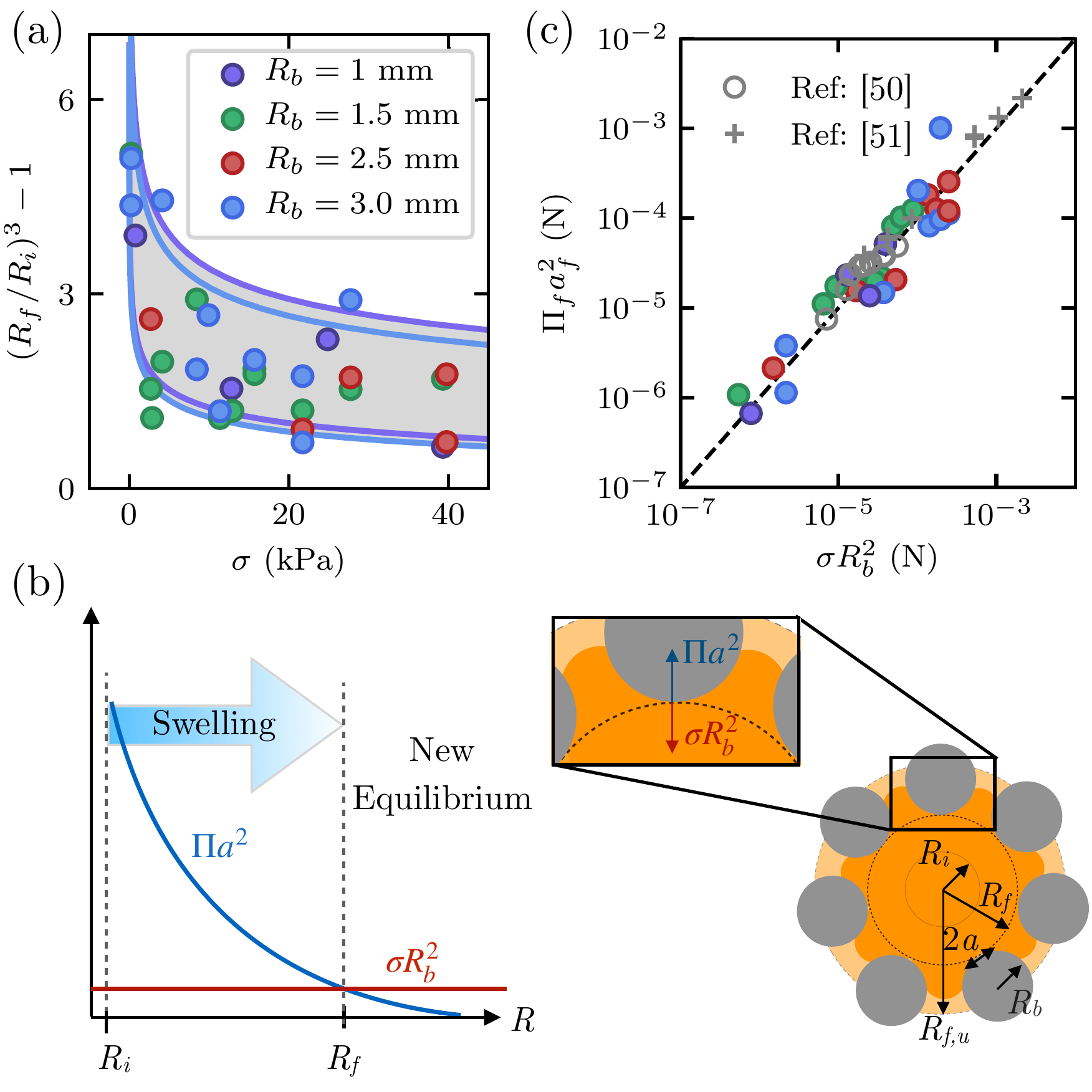}}\\
      \caption{Hydrogel swelling is determined by the competition between osmotic swelling and local confinement. (a) Fractional change in hydrogel volume $\sim R^{3}$ decreases with increasing stress $\sigma$, in media with different mean bead radii $R_{b}$. Curves show the prediction of Eq. \ref{eq_balance}; colors show $R_{b}$, and different sets of curves show $\pm~1$ standard deviation in $R_i$ from the measured mean. (b) Net osmotic swelling force (blue curve) decreases as the hydrogel swells and is eventually balanced by the confining force transmitted by beads (red curve). Right schematic shows a hydrogel of initial radius $R_{i}$ (inner circle), hindered swollen radius $R_{f}$ (dark orange with dashed circle), and equilibrium unconfined swollen radius $R_{f,u}$ (light orange) surrounded by beads (gray). Inset illustrates the force exerted by beads (red) and the force exerted by the swelling hydrogel at the contact (blue). (c) Hydrogel swelling in confinement is described by the balance between the net osmotic swelling force $\sim\Pi_{f} a_f^2$ and the confining force transmitted by the medium $\sim\sigma R_{b}^{2}$, each computed from independent measurements, as shown by the dashed line. Gray points show measurements of water absorption by hydrogels in soil from \textit{(50, 51)}.}\label{Fig3}
 \end{figure*}

We further explore the role of confining stress by performing experiments with varying $\sigma$, ranging from $0.2$ to $40$ kPa. Consistent with the observations in Figs. \ref{Fig1}-\ref{Fig2}, we find that swelling is highly sensitive to confinement: the final hydrogel size decreases strongly with increasing $\sigma$. We quantify this decrease by measuring the dependence of $R_{f}$, which directly characterizes the size of the region of hindered hydrogel swelling in contact with the surrounding beads, with $\sigma$. As shown in Fig. \ref{Fig3}a (blue), $R_{f}$ continually decreases with increasing $\sigma$, eventually approaching the limit of $R_{f}\approx R_{i}$. We find similar behavior for media with three smaller bead sizes $R_{b}=2.5$, $1.5$, and $1$ mm, indicated by the red, green, and purple symbols in Fig. \ref{Fig3}a, respectively. Each point represents a separate experiment done with a separate hydrogel, resulting in slight scatter in the data due to the slight experimental variability in $R_{i}$; within this scatter, the extent of hindered swelling does not appear to be strongly dependent on the bead size. Together, these results suggest that hydrogel swelling is generally hindered in granular media. \\

\subsection*{Determinants of hydrogel swelling in confinement}

What are the physics that describe the hindered swelling of hydrogels in granular media? To answer this question, we first consider the swelling of an unconfined hydrogel. When exposed to a good solvent, an uncharged hydrogel swells to promote mixing between its polymer chains and the solvent, as characterized by the mixing pressure $\Pi_\text{mix}$. This pressure is largest initially, when the hydrogel radius $R$ is small and the volume fraction of polymer in the hydrogel $\phi\sim R^{-3}$ is large; it then decreases as swelling progresses and $\phi$ decreases. Specifically, as established by polymer solution thermodynamics \cite{flory1953,tanaka1978,hirotsu1994,quesada2011,rubistein2003,fernandez2002,hashmi2009,sakai_2020,li2012},
\begin{equation}
    \begin{aligned}
    \Pi_\text{mix}=-\frac{k_{\mathrm{B}} T}{\alpha^{3}}\left[\phi +\ln(1-\phi) +\chi \phi^{2}\right] 
    \label{eq_mix}
    \end{aligned}
\end{equation}
\noindent where $k_{\rm{B}}$ is the Boltzmann constant, $T$ is temperature, $\alpha$ is the effective diameter of a solvent molecule, and $\chi$ is the Flory-Huggins polymer-solvent interaction parameter. This mixing pressure is resisted by deformation of the hydrogel polymer network, as characterized by the elastic pressure $\Pi_\text{el}$. As established by affine network elasticity \cite{flory1953,tanaka1978,hirotsu1994,quesada2011,rubistein2003,fernandez2002,hashmi2009,sakai_2020},
\begin{equation}
\begin{aligned}
\Pi_\text{el}=\frac{k_{\mathrm{B}} T N_{c}}{V_{0}}\left[\frac{\phi}{2 \phi_{0}}-\left(\frac{\phi}{\phi_{0}}\right)^{1 / 3}\right]
\label{eq_elastic}
\end{aligned}
\end{equation}
where $\phi_0$ and $V_{0}$ are the polymer volume fraction and hydrogel volume in a reference state---which for polyacrylamide is typically taken to be the preparation state in a good solvent with $\phi_{0}\ll1$ \cite{tanaka1978,hirotsu1994,quesada2011}---and $N_{c}$ is the number of polymer chains in the hydrogel network. The net osmotic swelling pressure of the hydrogel, $\Pi$, is then given by Flory-Rehner theory \cite{flory1953,tanaka1978,hirotsu1994,quesada2011,rubistein2003,fernandez2002,hashmi2009,sakai_2020}:
\begin{equation}
\begin{aligned}
\Pi=\Pi_\text{mix}+\Pi_\text{el}.
\label{eq_osmo}
\end{aligned}
\end{equation}
\noindent This osmotic swelling pressure is initially large, and solvent infiltrates the hydrogel network; however, as swelling progresses, $\Pi$ continually decreases and eventually reaches zero. At this equilibrium, $\Pi_\text{mix}$ and $\Pi_\text{el}$ balance each other, thereby defining the maximal swollen radius of an unconfined hydrogel, $R_{f,u}$. Thus, directly measuring $R_{f,u}$ for an unconfined, fully swollen hydrogel enables us to determine the parameters in Eqs. \ref{eq_mix} and \ref{eq_elastic} (\textit{Materials and Methods}).

Confinement in a granular medium changes this balance of stresses. Unlike the case of a spatially-uniform external stress, such as an imposed osmotic or hydrostatic pressure \cite{saunders,lietor}, a hydrogel swelling in a granular medium experiences a spatially non-uniform stress, \textcolor{black}{only at the hydrogel-bead contacts}. In particular, hydrogel swelling is resisted at regions in contact with the surrounding beads by both its own elastic pressure as well as the local confining stress transmitted by the beads. To understand this hindered swelling, we analyze the forces at the individual contacts, building on previous work investigating hydrogel swelling in the distinct case of planar confinement \cite{cai2011}. The force exerted by the swelling hydrogel at each contact region $\sim\Pi a^{2}$, where $a$ is the radius of this contact, as shown by the blue curve and blue arrow in Fig. \ref{Fig3}b\textcolor{black}{; here, both the osmotic swelling pressure $\Pi$ and the hydrogel-bead contact radius $a$ change as hydrogel swelling progresses}. This force is opposed by the force exerted by the surrounding beads. \textcolor{black}{While force transmission in disordered granular packings is known to be heterogeneous, previous work has shown that the distribution of contact forces is exponentially bounded above the mean \cite{mueth1998force}; thus,} for simplicity, we make the mean-field assumption that the confining stress $\sigma$ is distributed evenly through all the $\approx L^{2}/\pi R_{b}^{2}$ beads in a given $L\times L$ horizontal cross-section across the packing \cite{holtzman2012capillary}. The confining force $\sigma L^2$ transmitted per bead is then $\approx\sigma L^2/\left(L^{2}/\pi R_{b}^{2}\right)\sim\sigma R_{b}^{2}$, shown by the red arrow in the inset to Fig. \ref{Fig3}b. Hence, hydrogel swelling is hindered, reaching a smaller final inscribed radius $R_{f}$ given by the force balance\begin{equation}
\Pi_{f}a_{f}^{2} = \sigma R_{b}^2
\label{eq_balance}
\end{equation}
as schematized by the dashed line in Fig. \ref{Fig3}b. Here $a_{f}$ and $\Pi_{f}$ are the bead-hydrogel contact radius and osmotic swelling pressure, respectively, at this final state of swelling; $\Pi_{f}$ is given by evaluating Eq. \ref{eq_osmo} at $R=R_{f}$. Inter-bead friction plays a negligible role in this analysis, as described further in the \textit{Materials and Methods}.

To quantitatively test this prediction, we determine both $\Pi_{f}a_{f}^{2}$ and $\sigma R_{b}^2$ independently for all the measurements of $R_{f}$ shown in Fig. \ref{Fig3}a. We evaluate $\Pi_{f}$ using Eq. \ref{eq_osmo} with $R=R_{f}$; we estimate $\phi$ by measuring the size of a deswollen hydrogel placed in a poor solvent, and use measurements of hydrogel stiffness at various states of swelling to independently determine the parameters in Eqs. \ref{eq_mix} and \ref{eq_elastic} (\textit{Materials and Methods}). Motivated by previous work \cite{Mukhopadhyay,Nordstrom}, we use Hertzian contact mechanics \cite{johnson1987} to evaluate the contact radius $a$ for a given hydrogel radius $R$:
\begin{equation}
a^3 = \frac{3\sigma\pi R_b^2 \bar{R}}{4E^{*}}
\label{eq_a}
\end{equation}
where $1/\bar{R}\equiv1/R+1/R_{b}$ and $E^{*}$ is a radius-dependent effective hydrogel Young's modulus. We obtain $E^{*}$ directly using a power-law fit to normal force-indentation measurements of a hydrogel swollen in the refractive index-matched solvent to varying values of $R$ (\textit{Materials and Methods}). Combining all of these measurements provides a direct test of Eq. \ref{eq_balance}, for all of our experiments performed using different confining stresses $\sigma$ and different bead radii $R_{b}$. All the measurements of hydrogel size are well described by our analysis of the hydrogel-bead forces, as shown by the gray region in Fig. \ref{Fig3}a. Specifically, the curves of different colors correspond to the fractional volume change predicted by Eq. \ref{eq_balance} for different bead sizes, with the upper and lower sets of curves representing the theoretical prediction corresponding to $\pm~1$ standard deviation in the measured values of $R_i$ from the mean; all the data fall within the region between the curves, shown in gray, indicating that all the measurements can be captured by our theory. Furthermore, despite the scatter in the data and the mean-field assumption made in the theory, all of our measurements collapse onto a single linear relation between $\Pi_{f}a_{f}^{2}$ and $\sigma R_{b}^{2}$ that remarkably extends over nearly four orders of magnitude, as shown in Fig. \ref{Fig3}c. The close agreement between the theory and experiment thus indicates that the extent of hindered hydrogel swelling is determined by the competition between the force due to osmotic swelling and the local confining force transmitted by the beads of the medium.\\

\subsection*{Interplay between swelling and restructuring of the granular medium} 

\begin{figure*}
{\includegraphics[width=\textwidth]{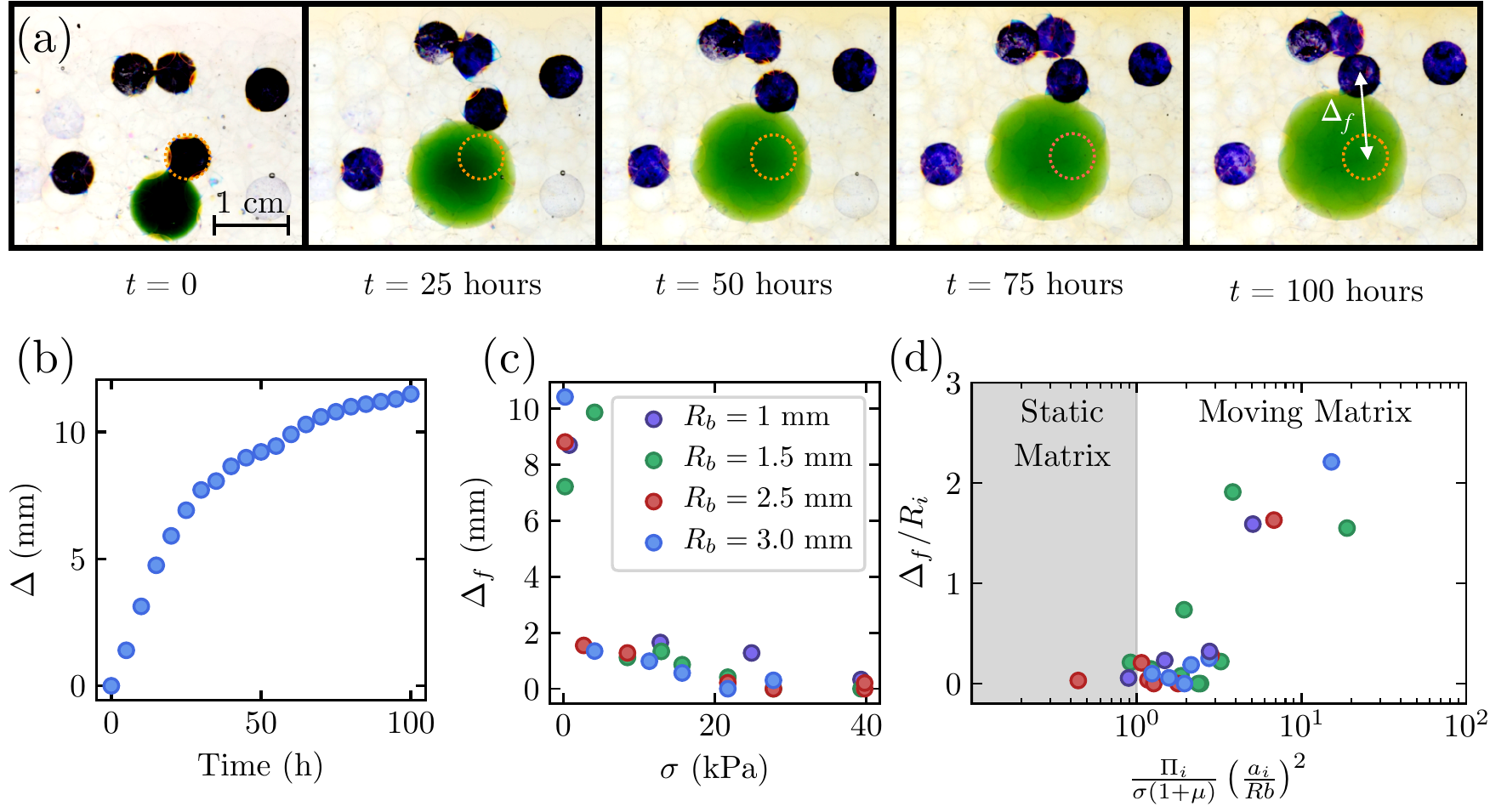}}
    \caption{Unbalanced hydrogel swelling restructures the surrounding medium. (a) Images show a swelling hydrogel (green) restructuring the surrounding medium, as indicated by the motion of dyed tracer beads (blue), for an experiment with small confining stress $\sigma=0.2$ kPa. The dashed orange line indicates the initial location of the bead showing the maximal displacement $\Delta$. (b) Maximal bead displacement as a function of time for an experiment with small confining stress $\sigma=0.2$ kPa. (c) Measurements of the plateau value of the maximal bead displacement $\Delta_{f}$ as a function of confining stress $\sigma$, for experiments with hydrogels confined in media with different mean bead radii $R_{b}$. The medium is restructured by hydrogel swelling at low confining stress, but is restructured less as confining stress increases. (d) The onset of medium restructuring, quantified by the normalized bead displacement, arises when the swelling number $N_{s}\equiv \Pi_i\left(a_{i}/R_{b}\right)^{2}/\sigma(1+\mu)$ becomes sufficiently large.}
\label{Fig4}
  \end{figure*}

Not only does confinement in a granular medium alter hydrogel swelling, but swelling can also alter the medium in turn. For example, as shown in Movie S1 for the load-free case, the beads overlying the hydrogel are pushed outward as it swells, with the beads immediately adjacent to the hydrogel undergoing maximal displacement. To \textcolor{black}{characterize} this behavior, we randomly colorize a sparse number of beads of the medium to act as tracers of medium restructuring, as shown in dark blue in Fig. \ref{Fig4}a, \textcolor{black}{and measure the magnitude of their displacement over the duration of each experiment. We denote the maximal measured displacement magnitude among this representative set of tracked beads by $\Delta$; for a static matrix, $\Delta\approx0$, while as the matrix is increasingly restructured, $\Delta$ increases above zero as the component of bead displacements orthogonal to the imaging direction increases. For the case of a low confining stress $\sigma\approx0.2$ kPa,} restructuring of the medium increases over time and $\Delta$ eventually plateaus to a value $\Delta_{f}$, concomitant with the hydrogel swelling, as shown in Fig. \ref{Fig4}b and Movie S4. As shown in Fig. \ref{Fig4}c, the maximal amount of medium restructuring \textcolor{black}{first decreases sharply, then gradually,} with increasing $\sigma$ up to $\sim20$ kPa; above this threshold, the medium is static and $\Delta_{f}\approx0$, as shown in Movie S5. As in Fig. \ref{Fig3}a, each point of Fig. \ref{Fig4}c represents a separate experiment done with a separate hydrogel, resulting in slight scatter in the data due to slight experimental variability in $R_{i}$.


To understand this behavior, we analyze the forces on a bead immediately adjacent to the hydrogel, inspired by previous work on fluid injection into a granular medium \cite{macminn2015, auton2019}. At any given time during hydrogel swelling, the net force on the bead can be approximated by $\Pi\pi a^{2}-\sigma\pi R_{b}^{2}$ within our mean-field assumption of even stress distribution across the packing; the first term represents the force exerted by the swelling hydrogel, while the second term represents the force exerted by opposing beads, as schematized in Fig. \ref{Fig3}b. For the beads surrounding a hydrogel to rearrange, this net force---which is directed normal to the inter-bead contacts---must exceed the limiting value of the inter-bead friction that resists rearrangement, $\approx\mu\sigma\pi R_{b}^{2}$ \cite{holtzman2012capillary}. Evaluating this net force at the initiation of hydrogel swelling---when the net force exerted on the beads by the hydrogel is maximal, as schematized in Fig. \ref{Fig3}b---thus yields a criterion for the onset of medium restructuring:
\begin{equation}
\Pi_i a_{i}^{2} -\sigma R_{b}^2 \gtrsim \mu \sigma R_b^2\textrm{,}~~\textrm{or}~N_{s}\equiv\frac{\Pi_i }{\sigma \left(1+\mu\right)}\left(\frac{a_{i}}{R_{b}}\right)^{2}\gtrsim1
\label{eq_pressures}
\end{equation}
\noindent where the subscript $i$ denotes the initiation of swelling, $\mu = 0.15$ is the coefficient of static friction between glass surfaces in an aqueous solvent as measured previously \cite{li2005comparison}, and $N_{s}$ is a dimensionless parameter we call the ``swelling number''. To test this idea, we evaluate $N_{s}$ using Eqs. \ref{eq_mix}-\ref{eq_osmo} and \ref{eq_a} for all of our experiments performed using different $\sigma$ and $R_{b}$. Consistent with our expectation, we find that all the measurements of $\Delta$ collapse into two regimes, \textcolor{black}{as shown in Fig. \ref{Fig4}d}: for $N_{s}\lesssim1$, $\Delta\approx0$---indicating that hydrogel swelling is insufficient to restructure the medium---while for sufficiently large $N_{s}\gtrsim1$, $\Delta$ monotonically increases, indicating that hydrogel swelling increasingly restructures the surrounding medium. Therefore, not only is hydrogel swelling impacted by confinement in a granular medium, but the medium in turn can be restructured by unbalanced hydrogel swelling.

 \newpage

 \section*{Discussion}
 By directly visualizing the swelling of hydrogels confined within 3D granular media, we have revealed the coupled dynamics of hydrogel swelling and medium restructuring.  \textcolor{black}{Unlike in previous investigations of hydrogel swelling under uniform stress, the inherent granularity of the medium imposes a spatially non-uniform stress on the hydrogel, only at the hydrogel-grain contacts. As a result, hydrogel swelling is hindered and spatially non-uniform.} Our \textcolor{black}{theoretical framework explicitly considers this non-uniformity by combining Flory-Rehner theory and Hertzian contact mechanics to compute the balance of forces at the hydrogel-grain contacts. We find good agreement between the prediction of this theory and our measurements of} the extent of hindered hydrogel swelling, \textcolor{black}{indicating that hydrogel swelling is controlled} by the competition between the osmotic swelling force and the local confining force transmitted by surrounding grains, as quantified by Eq. \ref{eq_balance}.

 Our analysis also indicates that the medium begins to restructure when the dimensionless parameter $N_{s}$ exceeds a threshold value $\sim1$, as quantified by Eq. \ref{eq_pressures}. Thus, the framework developed here may be used to predict the extent of hydrogel swelling and onset of medium restructuring given the physico-chemical properties of a hydrogel (as quantified by the parameters in Eqs. \ref{eq_mix}-\ref{eq_elastic}), the grain size $R_{b}$ and friction coefficient $\mu$ characterizing the medium, and the confining stress $\sigma$. We therefore expect that our findings will find broad use in diverse applications of hydrogels in granular media. \\

 \noindent\textbf{Directions for future work.} Given the simplifications made in our analysis—--that the hydrogel swelling force can be predicted by modifying the Flory-Rehner framework, that the force transmission is uniform across beads, and that the boundaries of the packing do not influence the experiments--—the close agreement between our theoretical predictions and our measurements suggests that our analysis provides a useful first step toward elucidating the essential physics governing hydrogel swelling in granular media. However, employing more sophisticated models of hydrogel swelling, heterogeneous force transmission in the granular matrix, and boundary effects will be an important direction for future research. Moreover, because our work represents a first step toward fully unraveling the physics underlying hydrogel-grain interactions, it necessarily involves some constraints. For example, to accomplish the experimental visualization, we use media that are limited in size, spanning $\sim7$ to $22$ grains across. Thus, boundary effects---such as friction between the glass beads and the container walls---could play a role in our experiments. These effects are known to depend on a complex interplay between the cross-sectional width of the packing, the height of the packing, grain size, friction, and packing history; estimating the magnitude of these effects is still an active area of research. However, recent experiments on confined granular packings suggest that for our bead sizes and packing widths, boundary friction does not appreciably alter the stress transmitted through the packing \cite{janssen}. Furthermore, we do not test grain/hydrogel size ratios $R_{b}/R_{i}<0.2$. However, we expect that the extent of hindered hydrogel swelling is not strongly influenced by the grain size: using the theory given in Eq. \ref{eq_balance}, we estimate that a hundredfold decrease in $R_{b}/R_{i}$ only alters the extent of hindered swelling by $\sim0.1$, as shown in \textcolor{black}{Fig. S1}. This expectation is consistent with the data shown in Fig. \ref{Fig3}a, which do not show a strong grain size dependence. Investigating hydrogel swelling in larger granular media, as well as over a broader range of hydrogel and grain sizes, will be a useful extension of our work. 

 Our experiments and analysis only focus on the region of hindered hydrogel swelling due to contacts with surrounding grains, as quantified by the inscribed circle of radius $R_{f}$, describing the growth of this region using a mean-field treatment of force transmission through the granular medium. Strikingly, this simple picture closely captures the extent of hindered swelling and the onset of medium restructuring, as shown in Figs. \ref{Fig3}c and \ref{Fig4}d, respectively. However, it does not consider hydrogel swelling in between the regions of grain contact, which we anticipate will increasingly contribute to the overall swollen hydrogel volume as $R_{b}/R_{i}$ increases; nor do we consider heterogeneity in force transmission, which could further cause heterogeneity in hydrogel swelling, or spatial variations in the full displacement profile of beads in the medium during restructuring. Exploring these complexities will also be an interesting direction for future work. 
 
   The analysis underlying Fig. \ref{Fig4}d focuses on determining the conditions under which the granular medium plastically rearranges, or yields. Specifically, beads rearrange when the net force on a bead $F_{net}\approx \Pi\pi a^2-\sigma\pi R_{b}^2$ exceeds the limiting value of the static friction at the onset of sliding $\approx \mu \sigma \pi R_{b}^2$. \textcolor{black}{While this analysis only considers pairwise interactions between beads, rearrangements may collectively involve several beads; the limiting value of the static friction may then need to be multiplied by a factor greater than one to incorporate multi-bead rearrangements. This complexity could explain why the predicted transition in \ref{Fig4}d occurs for $N_{s}$ slightly larger than one; incorporating this effect in a more sophisticated analysis of the medium restructuring would be a valuable direction for future work. More generally, the limiting value of the static friction} can be expressed in terms of the ``frictional yield stress" of the granular packing, $\tau_{f,y}=\mu\sigma$, which has been well-characterized for frictional packings at low deformation rates \cite{friction1}; that is, the granular medium is restructured by hydrogel swelling when $F_{net}$ exceeds the force on a bead required to overcome the frictional yield stress, $\tau_{f,y}\pi R_{b}^2$. We expect that this analysis can be generalized to packings with other interactions (\textit{e.g.}, inter-bead attraction) using the generalized yield stress $\tau_y$. In particular, we expect that the medium is restructured when $F_{net}\geq\tau_{y}\pi R_{b}^2$, where the yield stress $\tau_y$ becomes non-zero at and gradually increases above the jamming transition \cite{jamming1}.

 Finally, while this analysis focuses on the swelling of a single hydrogel in a granular medium, interactions between multiple hydrogels may also influence swelling, depending on the applied load and the volume fraction of the added hydrogels. For small imposed stress, we expect that grain rearrangements cause deformations in the medium to be localized to the surface of each swelling hydrogel; in this case, the swelling of one hydrogel likely does not influence the swelling of another. However, under a larger imposed stress, we expect that deformations may persist over larger length scales, due to the stronger coupling between the grains of the medium; thus, the swelling of one hydrogel could hinder the swelling of a neighboring hydrogel, mediated by the medium in between them, if the inter-hydrogel spacing is sufficiently small. For even larger imposed stresses under which the medium is not deformed by hydrogel swelling, we again expect that hydrogels swell independently of each other.\\

\noindent\textbf{Implications for applications of hydrogels.} In the context of agriculture, our results help rationalize recent laboratory \cite{lejcus2018swelling, misiewicz2019} and field \cite{huettermann2009application, narjary2012water} observations that hydrogels in the deeper layers of soil---which are subjected to a larger confining stress---absorb less water and swell less than hydrogels in upper layers. Indeed, the confining stress experienced by a hydrogel buried just $\sim1$ m below the soil surface can exceed $\sim10$ kPa, comparable to the swelling pressure of many commercially-used hydrogels, possibly leading to hindered swelling; this effect may underlie the large variability in hydrogel performance observed in the field. To test this idea, we examine the data for the mass of water absorbed by hydrogels in unconsolidated soil under load reported in  \cite{lejcus2018swelling, misiewicz2019} within the theoretical framework developed here (\textit{Materials and Methods}). Notably, despite the additional complexity inherent in these prior experiments, the reported data can be closely described by the force balance given by Eq. \ref{eq_balance}, as shown by the gray symbols in Fig. \ref{Fig3}c---indicating that our description of hindered swelling is relevant to the use of hydrogels in agriculture. We expect that different hydrogels with values of $\Pi_{f}$ tuned to the confining stresses and depths they are meant to be used at, as quantified by Eq. \ref{eq_balance}, may yield better absorption of water, potentially helping to address growing demands for food and water.

Another emerging application of hydrogels is in oil recovery: dry hydrogel particles injected into a reservoir have potential to swell and occlude high permeability pores, potentially enhancing oil recovery from lower permeability regions \cite{oconnell2019,yao2012,yao2015,bai2007}. Our results suggest that the degree to which the hydrogel swells---and hence, its permeability and ability to redirect flow \cite{li2015universal}---will depend sensitively on its physico-chemical properties as well as those of the reservoir rock. 

Finally, hydrogels are increasingly finding use in diverse other applications---e.g. as additives in construction materials \cite{krafcik2017improved}, as force sensors in packed tissues \cite{dolega2017cell,lee2019dispersible}, and as additives to membrane filters \cite{kapur1996}---that frequently involve their confinement in tight and tortuous spaces. These applications typically require hydrogel behavior to be predictable and controllable. Thus, the principles established here could be used more broadly to describe hydrogel swelling in diverse settings.

\newpage
\section*{Materials and Methods}
\noindent \textbf{Details of materials used.} The beads used to make 3D disordered granular media are borosilicate beads of mean radii $R_b=$ 1, 1.5, 2.5, or 3 mm obtained from \textit{Sigma-Aldrich}. The beads are packed in a clear leak-proof acrylic box constructed in-house, with an overlying piston constructed using a square piece of steel, 1 cm in thickness, with an embedded vertical steel rod that holds a prescribed number of metal plates of known mass. The mass $m$ used to impose a confining stress thus includes the combined mass of the piston and any other components used. The NH$_{\rm{4}}$SCN used for refractive index-matching is also obtained from \textit{Sigma-Aldrich}. The polyacrylamide hydrogel particles are water gel beads obtained from \textit{Jangostor}. \textcolor{black}{As noted in the main text, in each experiment, we overfill the packing with the solvent so the liquid surface is much higher than the position of the piston. While the liquid level falls slightly due to evaporation, the variation in the liquid level over the entire duration of the experiment is at most a centimeter---corresponding to a variation in hydrostatic stress less than $\sim100$ Pa, much smaller than the confining stress $\sigma$, which varies between $\sim200$ and $40,000$ Pa. Thus, we neglect evaporation-induced variations in the liquid level from our analysis.}\\

\noindent\noindent\textbf{Applicability of Hertzian contact mechanics.} Hertzian contact mechanics provides a simplified first step toward estimating the contact stresses, and thereby identifying the essential physics underlying hindered hydrogel swelling in confinement. Two key features of our experiments suggest that Hertzian contact mechanics provides a relevant first approximation to the contact stresses.

First, Hertzian contact mechanics assumes small strains that are within the limit of linear elasticity. Using our visualization, we estimate the maximal local strain as $\approx\left(R_{f,u}-R_{f}\right)/R_{f,u}\approx50\%$. While this maximal strain is considerable, it is nevertheless well within the regime of linear elasticity of the hydrogels we use, which exhibit linear elastic behavior to strains as large as $\sim100\%$ \cite{storm}. Thus, while the local strain associated with differences in hydrogel swelling is considerable, it is well within the range of linear elasticity, suggesting that Hertzian contact mechanics may provide a relevant first approximation to the contact stresses.

Second, our application of Hertzian contact mechanics at each contact is specifically using the Hertzian prediction for two contacting spherical elastic bodies of dissimilar sizes \cite{johnson1987}; in our case, these represent the hydrogel and an adjacent bead. Thus, our use of Hertzian contact mechanics explicitly incorporates the spherical geometry and sizes of both the hydrogel and the beads. However, a key assumption of Hertzian contact mechanics is that the radii of curvature of the contacting bodies, $R_f$ and $R_b$, are large compared with the radius of the circular contact region, $a$. In our experiments, $a/R_{f}\approx0.04$ and $a/R_{b}\approx0.2$ at the lowest applied loads tested, for which the Hertzian assumption is satisfied; thus, our simple theory likely provides a good first approximation of the contact stresses and helps identify the essential physics governing hindered hydrogel swelling in a granular medium. As the applied load is increased, however, $a/R_{f}$ and $a/R_{b}$ increase to $\approx0.6$ and $2$, indicating that the Hertzian assumption begins to break down and a more sophisticated non-linear theory will be required. This breakdown of the Hertzian assumption possibly explains the increasing deviation of our measurements from our simple theory at the largest confining stress $\sigma$ in Fig. \ref{Fig3}c.

Thus, given that the maximal strains are well within the linear elastic regime of the hydrogel, and that the contact sizes are small over a broad range of the loads tested, Hertzian contact mechanics likely provides a relevant first approximation to the contact stresses. However, a more sophisticated, non-linear theory suitable for large deformations could provide more accurate results, particularly at large values of the applied load; employing such non-linear theory to model hydrogel swelling in granular confinement will be a useful direction for future work.

\textcolor{black}{We also note that a key assumption of Hertzian theory is that we do not have any adhesion between the gel and the glass beads. Indeed, during removal of the hydrogel from the granular medium at the end of each experiment, we do not observe any adhesion between the gel and the glass beads---supporting this assumption. This assumption is further corroborated by previous measurements of contact forces between glass beads and polyacrylamide hydrogels, which can be described well using Hertzian contact mechanics \cite{simic2020,long2011,huth2019}. However, for cases in which hydrogel-bead adhesion is non-negligible, Eq. \ref{eq_a} would need to be replaced by the prediction from a theory that explicitly incorporates adhesion, such as Johnson-Kendall-Roberts (JKR) theory.}\\

\noindent\textbf{Possible influence of elastocapillarity.} Although the hydrogel deformations (\textit{e.g.}, fingering into the pore space between beads) may appear reminiscent of fluid wetting in some cases, we do not consider elastocapillary interactions for three key reasons.

First, our experiments only involve a single fluid phase---the aqueous NH$_{\rm{4}}$SCN solution---with the entire granular medium and the hydrogel submerged in this fluid phase. Thus, the hydrogel is not exposed to any immiscible fluid-fluid interfaces that could exert capillary forces.

Second, previous measurements \cite{king} indicate that the swollen hydrogel-aqueous interface itself is characterized by a vanishingly small interfacial tension $\gamma$: the strength of any possible capillary forces arising from the interface between the fluid-swollen hydrogel and the surrounding fluid is zero, to within measurement error, for acrylamide-based gels similar to those used in our experiments. The elastocapillary length scale over which capillary forces may deform the hydrogel $\sim\gamma/E$ thus becomes negligible. Indeed, for this length scale to be comparable to the deformation length scales observed in our experiments (larger than $\sim1$ mm), the interfacial tension would need to exceed $10^{4}$ mN/m, over a hundred times larger than the interfacial tension between air and water. 

Third, to further explore the possibility that the swollen hydrogels may exhibit wetting on the glass bead surfaces, we place a hydrogel bead swollen in the aqueous NH$_{\rm{4}}$SCN solution on a flat glass surface with similar surface properties as the glass beads comprising the granular media. Over a time scale of 100 h, comparable to the time scale of our experiments of hydrogel swelling in granular media, we observe no appreciable deformation of the hydrogel at the region of contact with the glass, as shown in \textcolor{black}{Fig. S2}. The hydrogel shrinks slightly due to drying, and a water meniscus forms at the contact region; however, we observe no appreciable wetting on the $\sim1$ mm length scale characterizing swelling in the granular media, confirming that the hydrogel behavior is dominated by elasticity, not capillarity. \\

\noindent\textbf{Measurements of hydrogel elasticity.}  To determine the effective hydrogel Young's modulus $E^{*}\equiv E/(1-\nu^{2})$, where $E$ is the Young's modulus and $\nu$ is Poisson's ratio, we use normal force-indentation measurements \cite{johnson1987} performed using parallel plates in a rheometer (Anton-Paar MCR 301) on a hydrogel swollen in the refractive index-matched solvent to varying values of $R$. To assess possible time-dependence in the indentation measurements, we perform them at two different indentation rates. We obtain similar measurements in both cases, as shown in \textcolor{black}{Fig. S3}a; the dashed line shows a fit to the data obtained at the slower indentation speed, but also provides an excellent fit to the data obtained at the faster indentation speed. Thus, the hydrogel elastic properties do not appear to be strongly time dependent for this range of indentation speeds. 

 
We use these measurements, summarized in \textcolor{black}{Fig. S3}b, to calculate the hydrogel shear modulus $G$ via the relation $G=E/2(1+\nu)$, using $\nu\approx0.45$ as previously measured \cite{traeber2019polyacrylamide}. We compare the measured values to the prediction from network elasticity \cite{rubistein2003}, $G=k_\textrm{B}TN_{c}/V$, where $N_{c}$ is the number of polymer chains in a hydrogel particle of volume $V$, which yields $N_{c}\approx5\times10^{18}$. While this prediction does not capture the full $R$ dependence of the modulus measurements, it provides a first approximation to estimating the shear modulus; indeed, while we use the power-law $E^{*}\sim R^{-1.6}$ obtained using least-squares fitting to enable determination of the hydrogel-bead contact radius $a$ for any given hydrogel radius $R$, we find that the data can also be reasonably fit by the network elasticity prediction $E^{*}\sim R^{-3}$, as shown by the dashed line in \textcolor{black}{Fig. S3}b. Using this fit in Eqs. \ref{eq_balance}-\ref{eq_a} instead does not appreciably change our results; we still observe good agreement between our calculations and our theoretical prediction, as shown in \textcolor{black}{Fig. S4}. Thus, our central findings are robust to variations in the choice of the fitting function used to describe $E^{*}(R)$.

To further assess possible time-dependence in the indentation measurements, we compare this indentation time to the poroelastic time scale $\tau_{p}$ over which poroelastic effects, arising from the coupling between hydrogel deformation and fluid flow through the internal hydrogel mesh, equilibrate for our experiments in granular media. We estimate $\tau_{p}\approx(\mu L^2)/\left(E^{*}k\right)\sim 20$ min for a deformation over the granular length scale $L\sim1$ mm; here, $\mu\approx 1$ mPa-s is the solvent dynamic shear viscosity, $E^{*}\approx15$ kPa is the hydrogel Young’s modulus, and $k\approx 5\times 10^{-17}$ m$^{2}$ is the internal permeability of the hydrogel, estimated based on measurements performed on other hydrogels of a composition similar to ours \cite{oyen}. The time scale of the indentation experiments, $\sim 20$ min, is comparable to the poroelastic time scale, $\tau_{p}\sim20$ min, suggesting that they are sufficiently long to not be strongly influenced by time-dependent poroelastic effects.

As a final test of this expectation, we measure the elastic properties of a hydrogel over time scales comparable to those of our experiments of hydrogel swelling in granular confinement. Specifically, we use a flat plate to impose a fixed, constant indentation of $\delta=3$ mm on a fully swollen hydrogel immersed in a bath of the NH$_{\rm{4}}$SCN aqueous solution used as the solvent, and measure the normal force exerted on the plate over a duration of 12 h. As expected, the measured normal force decreases slightly over time, indicating a slight influence of poroelastic effects. However, over the experimental duration, this decrease is at most $10\%$ of the initially measured force, as shown in \textcolor{black}{Fig. S3}c, indicating that the corresponding effective Young’s modulus decreases by at most $10\%$. The hydrogel-bead contact radius calculated using Eq. \ref{eq_a}  thus changes by $\lesssim4\%$. These experiments therefore confirm our measurements of hydrogel elasticity are not strongly influenced by time-dependent poroelastic effects. \\

\noindent\textbf{Determination of parameters in Flory-Rehner theory.} For each measured hydrogel radius $R$, we estimate $\phi\approx\phi_{dry}\left(R_{dry}/R\right)^{3}$ using mass conservation, where $\phi_{dry}\approx1$ and $R_{dry}$ are the polymer volume fraction and radius of a completely deswollen hydrogel, respectively. To determine $R_{dry}$, we measure the radius $R_{ac}$ of a hydrogel particle equilibrated after two successive 24 h-long baths in acetone---a poor solvent for polyacrylamide. Under these conditions, $\phi\approx0.7$ \cite{hashmi2009}; we measure $R_{ac}=3.3$ mm and thus take $R_{dry}\approx3$ mm. Hence, $\phi\approx\left(3~\textrm{mm}/R\right)^{3}$ for the different values of $R$ we measure. 

Our measurements of hydrogel stiffness at various states of swelling yield an estimate of $N_{c}$, the number of polymer chains in a hydrogel particle of volume $V$, using network elasticity theory. We find $N_{c}\approx5\times10^{18}$. We independently determine $N_{c}$ using size measurements of an unconfined, fully-swollen hydrogel. Specifically, at equilibrium, an unconfined hydrogel swells to a radius $R_{f,u}$ at which $\Pi=0$. We estimate $R_{f,u}\approx1.2$ cm from direct measurements of hydrogel swelling, and use this value in Eqs. \ref{eq_mix}, \ref{eq_elastic}, and \ref{eq_osmo} to independently determine the value of $N_{c}$. Previous measurements of the Flory-Huggins polymer-solvent interaction parameter $\chi$ for polyacrylamide in water, a good solvent, yield $\chi=0.495$ \cite{day1981}. Because our aqueous solution of NH$_{\rm{4}}$SCN is a less good solvent for the hydrogel---which swells more in water than in NH$_{\rm{4}}$SCN---but is still not a bad solvent with $\chi>0.5$, we use $\chi=0.499$. Using the measured $R_{f,u}$ and estimated $\chi$ in Eqs. \ref{eq_mix}, \ref{eq_elastic}, and \ref{eq_osmo}, we again find $N_{c}\approx5\times10^{18}$ using $\alpha=4\times10^{-10}$ m and $\phi_{0}=0.6\%$, consistent with previous measurements \cite{fernandez2002,hashmi2009,hirotsu1994,sakai_2020,li2012,tanaka1978, flory1953,quesada2011}; the reference state volume $V_{0}$ is then directly given by $V_{dry}/\phi_{0}$, where $V_{dry}\equiv (4/3)\pi R_{dry}^{3}$. We thus use these measurements as fixed values in the Flory-Rehner analysis described in the main text, except for $E^{*}$, for which we explicitly incorporate the measured size dependence $E^{*}(R)$. 

To analyze measurements of hydrogel swelling in unconsolidated soil within our framework, we examine the previous measurements of mass of water absorbed under different amounts of load reported in \cite{lejcus2018swelling, misiewicz2019}. We estimate all values of the parameters in Eqs. \ref{eq_osmo} and \ref{eq_balance} using the experimental details provided in \cite{lejcus2018swelling, misiewicz2019}. The values of $\sigma$ are directly given in these references. To estimate $R_{b}$, we use the median value of the grain size distribution reported in \cite{misiewicz2019} for both coarse soil and sandy loam, $R_b \approx 0.6$ mm and $R_b \approx 0.12$ mm, respectively, and we use the same sandy loam grain size to analyze the similar measurements in sandy loam from \cite{lejcus2018swelling}. To determine $R_{f}$, we convert the reported values of the mass of water absorbed for a given hydrogel mass, denoted $S_e$ (g/g), via the mass conservation relation $R_f = R_{i} (S_{e}+1)^{1/3}$, with $R_{i} = 6 \times 10^{-4}$ m from both. We determine $R_{wet}$ similarly by using the reported value of $S_e$ with no load or confinement. The hydrogel Young's modulus is not provided in either reference; hence, to estimate $E^*$ and thereby determine $a_{f}$ in Eq. \ref{eq_a}, we assume that the hydrogel mechanical properties can be estimated by treating it as a semidilute polymer solution, as experimentally verified by others for many hydrogels \cite{degennes1979,vandersman2015,zrinyi1987,stubbe2002,cho2019crack,cho2019crackprl}, with $E^* \approx E_{f,u} (R_f/R_{f,u})^{-27/4}$, where we take $E_{f,u}\approx 2$ kPa to be the fully swollen unconfined Young's modulus consistent with typical moduli measured for acrylamide-based hydrogels \cite{caporizzo2015}. Finally, to estimate $\Pi_{f}$ using Eqs. \ref{eq_mix}, \ref{eq_elastic}, \ref{eq_osmo}, we use the same Flory-Rehner parameters as in our experiments, given that the hydrogels in \cite{lejcus2018swelling,misiewicz2019} are made of a similar polymer as those used in our work, with some values modified to better represent the differences in the experiments reported in \cite{lejcus2018swelling,misiewicz2019}. Given that the solvent used is pure water, we use $\chi = 0.495$ as previously measured \cite{day1981}. Given that the initial hydrogel radius reported for experiments at standard humidity and temperature/pressure conditions is $R_{i} = 6 \times 10^{-4}$ m, we take the completely dry size to be slightly smaller, $R_{dry} \approx 5 \times 10^{-4}$ m. Finally, given that the hydrogels in \cite{lejcus2018swelling,misiewicz2019} are made of a similar polymer as those used in our work, we expect a chain density in the preparation state $N_{c}/V_{0}=N_{c}\phi_{0}/V_{dry}$ comparable to that of our hydrogels; we thereby estimate $N_{c}\approx 10^{18}$ and $\phi_{0}\approx0.1\%$. Using all of these values, we compute $\Pi_{f}a_{f}^{2}$ and $\sigma R_{b}^{2}$ for all of the measurements reported in \cite{lejcus2018swelling,misiewicz2019}; the resulting values are plotted in Fig. \ref{Fig3}c as the gray circles and squares. Despite the complexity inherent in these previous experiments, the results are closely consistent with those from our experiments and with our theoretical prediction given by Eq. \ref{eq_balance}---suggesting that our framework is more generally applicable.\\

\noindent\textbf{Negligible role of inter-bead friction in determining the extent of hydrogel swelling.} Our analysis of hindered hydrogel swelling, given by Eq. \ref{eq_balance}, does not consider inter-bead friction. A friction force $F_{f}$ could, in principle, be included in this analysis. In this case, the final equilibrium state of hydrogel swelling would be described by the relation $\Pi_{f}\pi a_{f}^2-\sigma\pi R_{b}^2=F_f$; Eq. \ref{eq_balance} represents the $F_f\rightarrow0$ limit of this expression, for simplicity. In practice, it is unclear what the magnitude of $F_f$ is, as this arises from residual unbalanced tangential forces between the beads of the packing. However, the limiting value of the static friction at the onset of inter-bead sliding $\approx \mu\sigma\pi R_{b}^2$ represents the maximal value of the inter-bead friction, and thus provides an upper bound for $F_f$. Substituting the coefficient of friction between glass surfaces in an aqueous solvent, $\mu= 0.15$, therefore yields a modified form of Eq. \ref{eq_balance} that incorporates the maximal value of inter-bead friction: $\Pi_{f}\pi a_{f}^2=1.15\sigma\pi R_{b}^2$. Importantly, this modified expression differs from Eq. \ref{eq_balance} by only $15\%$; given that our measurements of the force balance (shown in Fig. \ref{Fig3}c) span over three orders of magnitude, this error is negligible. We further show this negligible influence in \textcolor{black}{Fig. S5}, which shows Fig. \ref{Fig3}c with an additional curve representing the force balance incorporating the maximal value of the inter-bead friction; all the measurements agree with the combined predictions well. Thus, our central findings are unaffected whether or not friction is incorporated in the analysis.

\newpage\section*{Acknowledgments}
It is a pleasure to acknowledge the Stone lab for the use of the rheometer; Tapomoy Bhattacharjee, Chris Browne, and Sankaran Sundaresan for stimulating discussions; \textcolor{black}{and the anonymous reviewers for their helpful feedback}. This work was supported by the Grand Challenges Initiative of the \textcolor{black}{High Meadows} Environmental Institute, the ReMatch+ program at Princeton, the Lidow Senior Thesis fund, the Mary and Randall Hack Graduate Award of the \textcolor{black}{High Meadows} Environmental Institute to N.B.L., and in part by funding from the Princeton Center for Complex Materials, a Materials Research Science and Engineering Center supported by NSF grants DMR-1420541 and \textcolor{black}{DMR-2011750}.\\

\noindent\textbf{Author contributions:} All authors designed the experiments; J-F.L. and N.B.L. performed all experiments; J-F.L., N.B.L., and S.S.D. analyzed the data; S.S.D. and J-F.L. designed and performed the theoretical analysis; J-F.L., N.B.L., and S.S.D. discussed the results and implications and wrote the manuscript; S.S.D. designed and supervised the overall project.\\

\noindent\textbf{Data availability:} All data needed to evaluate the conclusions in the paper are present in the paper and/or the Supplementary Materials.\\

\noindent\textbf{Competing interests:} All authors declare that they have no competing interests.

\newpage

\newpage\section*{List of Supplementary Materials}
\noindent\textcolor{black}{\textbf{Figures S1-S5.} Analysis of the influence of bead size on hydrogel swelling;} measurements of hydrogel-glass interaction; \textcolor{black}{mechanical characterization of hydrogels;} force balance using different scaling for $E^{*}$; force balance including inter-bead friction.\\

\noindent\textbf{Movie S1.} An opaque 3D granular medium becomes transparent, and an embedded hydrogel (orange) becomes visible, when saturated with a refractive index-matching solvent.\\

\noindent\textbf{Movie S2.} Swelling of a hydrogel (orange) within a granular medium in the absence of an applied load ($m = 0$) and thus with a small confining stress $\sim$ 0.2 kPa. The hydrogel swells freely and retains its spherical shape.\\

\noindent\textbf{Movie S3.} Swelling of a hydrogel (orange) within a granular medium in the presence of
an applied load ($m \sim 4$ kg) and thus with a large confining stress $\sim$ 22 kPa. The hydrogel fingers into the surrounding pore space and has hindered swelling. The liquid level falls slightly due to evaporation, but always remains well above the bottom of the overlying piston. \\

\noindent\textbf{Movie S4.} Swelling of a hydrogel (green) within a granular medium in the absence of an applied load ($m = 0$) and thus with a small confining stress $\sim$ 0.2 kPa. Bead tracers are shown in blue. As the hydrogel swells, it rearranges the surrounding medium.\\

\noindent\textbf{Movie S5.} Swelling of a hydrogel (green) within a granular medium in the presence of an applied load ($m \sim 4$ kg) and thus with a large confining stress $\sim$ 22 kPa. Bead tracers
are shown in blue. As the hydrogel swells, it does not rearrange the surrounding
medium.

\end{document}